\documentclass[aps,prb,groupedaddress]{revtex4}

\usepackage{graphicx}
\usepackage{dcolumn}
\usepackage{epsfig}
\usepackage{bm}
\usepackage{amsfonts}
\usepackage{latexsym}

\newcommand{\be}{\begin{equation}}
\newcommand{\ee}{\end{equation}}
\newcommand{\bea}{\begin{eqnarray}}
\newcommand{\eea}{\end{eqnarray}}

\newcommand{\Tr}{{\rm Tr}}
\newcommand{\half}{ \frac{1}{2} }

\newcommand{\figwidth}{0.55\columnwidth}

\begin{document}


\title{Phase Diagram of the BCC $S=\frac{1}{2}$ Heisenberg Antiferromagnet\\
with First and Second Neighbor Exchange}


\author{J. Oitmaa}
\email[]{j.oitmaa@unsw.edu.au}
\affiliation{School of Physics,
The University of New South Wales,
Sydney, NSW 2052, Australia.}

\author{Weihong Zheng}
\email[]{w.zheng@unsw.edu.au}
\homepage[]{http://www.phys.unsw.edu.au/~zwh}
\affiliation{School of Physics,
The University of New South Wales,
Sydney, NSW 2052, Australia.}


%

\date{\today}

\begin{abstract}
We use linked-cluster series expansions, both at $T=0$ and high temperature, to
analyse the phase structure of the spin-$\half$ Heisenberg antiferromagnet with competing
first and second-neighbor interactions on the 3-dimensional body-centred-cubic
lattice. At zero temperature we find a first-order quantum phase transition at
$J_2/J_1 \simeq 0.705 \pm 0.005$ between AF$_1$ (Ne\'el) and AF$_2$ ordered
phases. The high temperature series yield quite accurate estimates of the bounding
critical line for the AF$_1$ phase, and an apparent critical line for the
AF$_2$ phase, with a bicritical point at $J_1/J_2\simeq 0.71$, $kT/J_1\simeq 0.34$.
The possibility that this latter transition is first-order cannot be excluded.
\end{abstract}




\maketitle

\section{\label{sec:intro}INTRODUCTION}

The occurrence of competing exchange interactions in magnetic materials
can give rise to a rich
variety of magnetic ordered states, and of phase transitions between them.
Studies of such phenomena, within the ``molecular" or ``mean-field"
approximation go back half a century\cite{sma66} or more. It is perhaps surprising
that open questions remain, but, at least for quantum models, this is the case.

We study in this paper, the spin-$\half$ Heisenberg antiferromagnet on the
body-centered-cubic (bcc) lattice, with first- and second-neighbor interactions.
The Hamiltonian is
\be
H=J_1 \sum_{\langle ij\rangle}{}^{{}^{(1)}} {\bf S}_i \cdot {\bf S}_j
+ J_2 \sum_{\langle ij\rangle}{}^{{}^{(2)}} {\bf S}_i \cdot {\bf S}_j
\ee
where the summations are over first and second-neighbor pairs
respectively and the exchange constants $J_1,J_2>0$.
For our purposes we need to divide the structure into four
interpenetrating  sublattices, which we denote $A_1$, $A_2$, $B_1$, $B_2$
and illustrate in Fig. 1(a). Each of these sublattices has a face-centered-cubic
(fcc) structure. The $J_1$ interactions couple the
$A$ and $B$ sublattices, while the interactions $J_2$ couple $A_1$ to $A_2$,
and $B_1$ to $B_2$ only.

The first question to ask concerns the ground state, at $T=0$.
For classical vector spins it is easy to show that this will be the
N\'eel or AF$_1$ phase for $J_2/J_1<2/3$, and the AF$_2$ phase for
$J_2/J_1>2/3$. In the AF$_1$ phase all A spin point in the direction
of an arbitrary unit vector $\hat{\bf{n}}$ while $B$ spins point in the opposite
direction $-\hat{\bf{n}}$. In the AF$_2$ phase each of the $A$ and $B$
sublattice is itself N\'eel ordered. Thus we may choose $A_1$ spins in
direction $\hat{\bf{n}}$, $A_2$ spins in direction $-\hat{\bf{n}}$,
$B_1$ spins in direction $\hat{\bf{n}}'$ and $B_2$ spins in direction
$-\hat{\bf{n}}'$, with $\hat{\bf{n}}$ and $\hat{\bf{n}}'$ completely independent.
In the quantum case these will not be exact eigenstates but, as
usual in quantum antiferromagnetism, will be modified by quantum fluctuations.
Furthermore in the AF$_2$ phase the direction $\hat{\bf{n}}'$ will be locked to
either $\pm\hat{\bf{n}}$, leading to a discrete 2-fold Ising symmetry added to the
O(3) spin symmetry.

This model has similarities to the so-called ``$J_1-J_2$ model" on the square
lattice, which has been much studied in recent years\cite{cha88,cap00,sus01}.
In that case there is now strong evidence for an intermediate ``spin-liquid"
phase between the N\'eel and AF$_2$ (the ``collinear phase"), with the
possibility of an even richer structure. Motivated by this similarity,
Schmidt {\it et al.}\cite{sch02} have studied the bcc model at $T=0$, using
exact diagonalizations and linear spin-wave theory. They find a direct first-order
quantum phase transition between  AF$_1$ and AF$_2$ phase at $J_2/J_1 \simeq 0.7$,
with no intermediate phase. This can be understood in terms of the diminishing effect
of quantum fluctuations in the 3-dimensional system.

In Section 2 we use linked-cluster series expansions at $T=0$ to study
ground state properties of the model. We obtain rather precise results
for both the ground state energy and order parameters, confirming
the scenario of Ref. [\onlinecite{sch02}] and obtaining a somewhat more
precise estimate of the quantum phase transition point.

A second interesting aspect of the model is the phase structure at
finite temperatures. There will be transition lines in the
($T$, $J_2/J_1$) plane where the low-temperature ordered phases
meet the high-temperature disordered phase, and the locations
and nature of these transition lines are not well known. The earliest
attempt to address these questions, beyond mean-field theory, was the 6th-order
high-temperature series work of Pirnie {\it et al.}\cite{pir66}.
They obtained series for the appropriate staggered susceptibilities for both AF$_1$ and AF$_2$
ordering for the $J_1-J_2$ bcc lattice (for general spin), and estimated critical
temperatures via Dlog Pad\'e approximants, in the usual way.
They were unable to obtain precise values for the critical exponent $\gamma$.
More recently Pan\cite{pan99} has computed an 8th-order staggered susceptibility
series for the N\'eel phase for the case $J_2=0$, and obtained the
critical temperature as $kT_c/J_1=1.384\pm 0.005$. (This is
actually a thermodynamic perturbation expansion, which can generate a high-temperature
series but can also used to study properties in the ordered phase).

In Section 3 we report on the derivation and analysis of new extended high
temperature series for this system. In particular we have computed staggered
susceptibilities for the full $J_1-J_2$ model to 10th order in the
AF$_1$ phase and 9th order in the AF$_2$ phase, adding 4 and 3 terms
respectively to the work of Ref. [\onlinecite{pir66}]. These longer series
allow reasonably precise estimates to be made of both critical
temperature and exponents. There is one caveat. There are arguments that the
AF$_2$ to paramagnetic transition is first-order. If that is indeed the case then our
series are presumably seeing a spinodal line within the ordered phase, and the actual
transition temperature is higher.

The arguments for a first-order transition come from two sources.
Firstly for the Ising version of this system both Monte Carlo
simulations\cite{ban79} and a combination of high and low-temperature
series\cite{vel83} support, fairly conclusively, the existence of
a ``fluctuation-induced" first-order transition. Secondly, a
renormalization group treatment\cite{ban79,shn80} of the appropriate
$n=6$ component Landau-Ginzburg-Wilson Hamiltonian for the
AF$_2$ transition has no stable fixed point, suggesting a first-order
transition. However this analysis is based on an $\epsilon$-expansion
about $d=4$ and, in our view, while persuasive, is not conclusive.

\begin{figure}[t]
  { \centering
    \includegraphics[width=\figwidth]{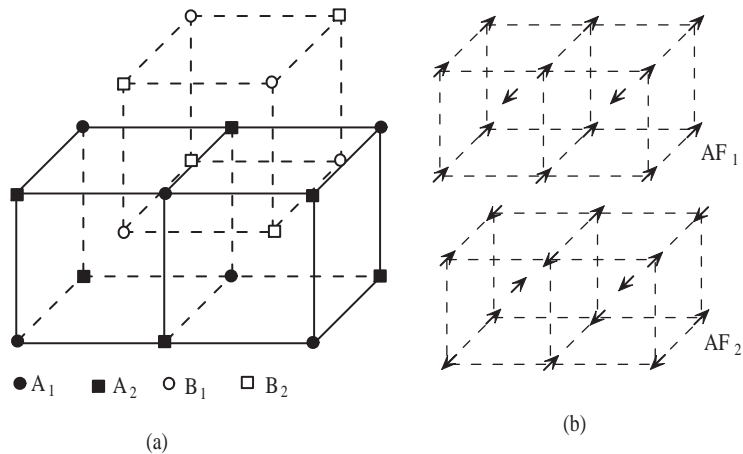}
    \caption{(a) The four sublattices of the bcc structure, (b) the AF$_1$ amd
    AF$_2$ ordered phases.
    \label{fig_1}}
  }
\end{figure} 

\begin{figure}[t]
  { \centering
    \includegraphics[width=\figwidth]{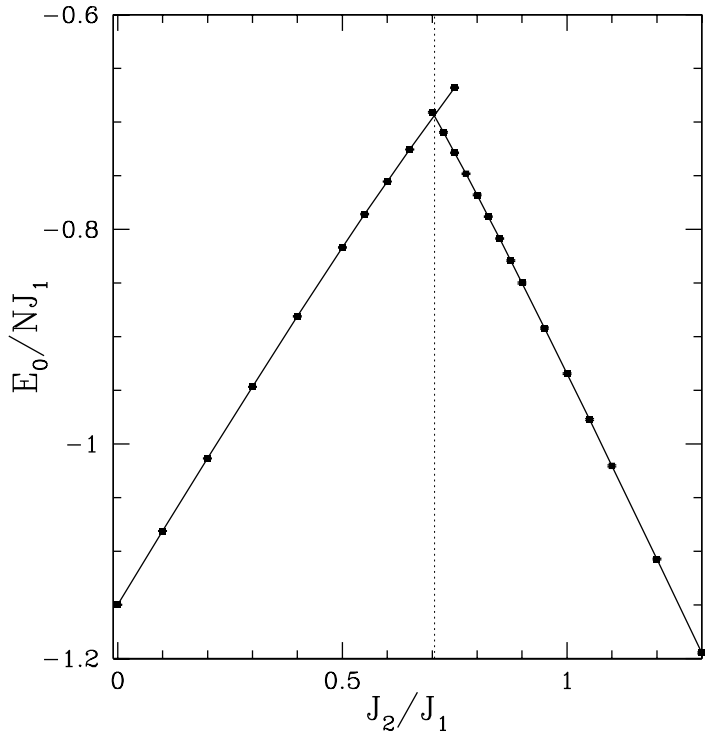}
    \vspace{0.5cm}
    \caption{Ground state energy versus $J_2/J_1$ from Ising expansions
    for the AF$_1$ and AF$_2$ phases. The crossing point at
    $0.705\pm 0.005$ identifies the first-order quantum phase transition.
    Uncertainties in the series extrapolation are no larger than the symbols.
    Lines joining symbols are guides to the eye.
    \label{fig_2}}
  }
\end{figure} 

\begin{figure}[t]
  { \centering
    \includegraphics[width=\figwidth]{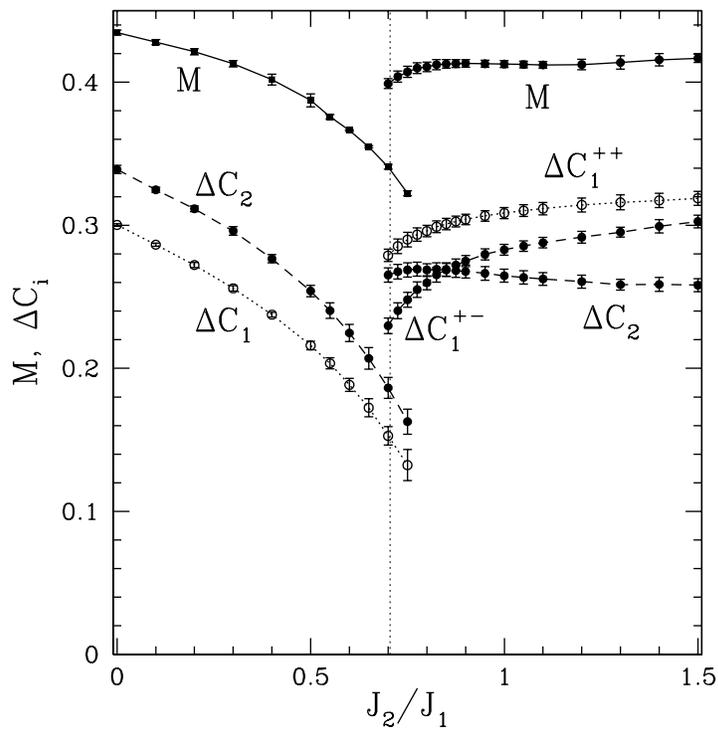}
    \vspace{0.5cm}
    \caption{Staggered magnetizations (order-parameters) and correlator discriminants
    (Eq. \ref{eq_dc}) versus $J_2/J_1$, in both the AF$_1$ and AF$_2$ phases.
    $\Delta C_1$, $\Delta C_2$ refer to first and second neighbour correlators respectively.
    In the AF$_2$ phase the superscripts $++$ and $+-$ refer to first neighbour
    correlators for sites with like spins and unlike spins.
    The thin vertical line denotes the transition. The lines joining symbols are
    guides to the eye.
    \label{fig_3}}
  }
\end{figure} 

\begin{figure}[t]
  { \centering
    \includegraphics[width=\figwidth]{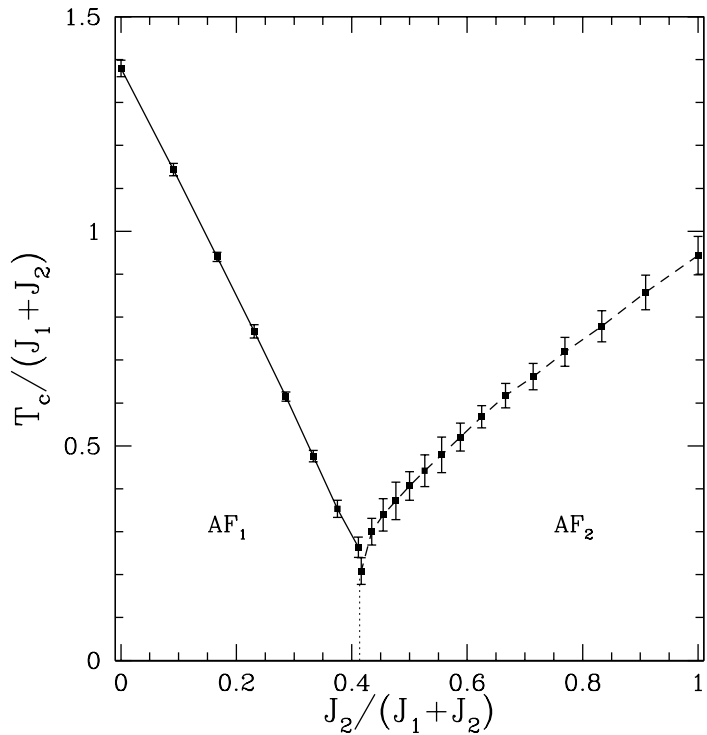}
    \vspace{0.5cm}
    \caption{Phase diagram of the BCC $J_1-J_2$ antiferromagnet. The solid line
    is the AF$_1$-paramagnetic critical line, with universal $n=3$ exponents.
    The dashed line is the AF$_2$-paramagnetic transition line, as determined from
    our series. The short vertical dotted line is the first-order AF$_1$-AF$_2$ transition
    line, extending from the bicritical point (or crtical end-point) to
    the quantum phase transition point on the $T=0$ axis.
    \label{fig_4}}
  }
\end{figure} 

\section{\label{sec2}Expansions at $T=0$}
Our approach is to use high-order linked-cluster expansions\cite{gel00}
where the Hamiltonian is written as
\be
H=H_0 + x V
\ee
where $H_0$ is the Ising Hamiltonian
\be
H_0 = J_1 \sum_{\langle ij\rangle}{}^{{}^{(1)}} S_i^z S_j^z +
J_2 \sum_{\langle ij\rangle}{}^{{}^{(2)}} S_i^z S_j^z
\ee
and
\be
V = \half J_1 \sum_{\langle ij\rangle}{}^{{}^{(1)}} ( S_i^+ S_j^- + S_i^- S_j^+)
+  \half J_2 \sum_{\langle ij\rangle}{}^{{}^{(2)}} ( S_i^+ S_j^- + S_i^- S_j^+)
\ee
is treated as a perturbation. Quantities are expanded as power series
in $x$, and then evaluated at the isotropic limit $x=1$, by Pad\'e
and differential approximants.
We have used this technique successfully in many previous studies of quantum
antiferromagnets\cite{zhe91,oit94}. In the present work we
are able to compute series through $x^{8}$  in both N\'eel and AF$_2$ region.

Figure 2 shows the ground state energy versus exchange ratio $J_2/J_1$,
the two branches corresponding to the two phases which can occur. The
results are very similar to those of Ref. [\onlinecite{sch02}] obtained from extrapolation of
exact diagonalization results. The two branches clearly cross, confirming that
the ground state transition between AF$_1$ and AF$_2$ phases is first-order.
Our estimated transition point is $J_2/J_1=0.705\pm 0.005$.

We have also computed series for two other physical quantities:
staggered magnetization in
each phase $M$, and the correlator discriminant
\be
\Delta C = \vert 3 \langle S_i^z S_j^z \rangle -
 \langle {\bf S}_i\cdot {\bf S}_j \rangle \vert \label{eq_dc}
\ee
which is a measure of the breaking of spin rotation symmetry.
These results are shown in Figure 3. We look first at the magnetization.
In the N\'eel phase this starts from $\sim 0.435$ and decreases monotonically to approx. 0.34
at the transition point. In the AF$_2$ phase the curve start from 0.424, the result for the
simple cubic lattice\cite{oit94}, and remains quite flat, with a downward curvature near
the transition point. Our results show a small upward curve near $J_2/J_1\simeq 1$, but this is
most likely a artifact. Our results are similar to, but considerably more precise than,
those of ref. [\onlinecite{sch02}].

Results for the correlators are new. Rather than show all correlators, we show in
Figure 3 only the quantities $\Delta C$ in Eq. \ref{eq_dc} for
first and second-neighbor pairs. A nonzero value indicates spontaneous breaking of spin
rotation symmetry. Of course our expansion starts from a symmetry-broken state,
and is thus biased. However an identical calculation for the
$J_1-J_2$ square lattice\cite{sus01} clearly shows $\Delta C\to 0$ at the
boundary between the N\'eel and disordered phase. The fact that this does not
happen here is clear and strong evidence that the N\'eel phase does not
vanish continuously, but undergoes a first-order transition to the AF$_2$ phase.
The second-neighbor discriminant $\Delta C_2$ in the AF$_2$ phase is particularly
interesting, showing a small but clearly discernable increase on approaching the
transition. This is unexpected.

In summary our $T=0$ series confirm the occurrence of two phase, AF$_1$ and AF$_2$ with
a first-order quantum phase transition at $J_2/J_1=0.705\pm 0.005$. We now turn to the finite
temperature phase structure.

\section{High Temperture Expansions\label{sec3}}
The standard approach to determining finite temperature critical points and
exponents for quantum (and classical) models is via high-temperature expansions.
For the bcc(1,2) antiferromagnet this was first done in Ref. [\onlinecite{pir66}].
One includes an appropriate staggered field $F$ in the Hamiltonian
\be
H' = H - F \sum_i \eta_i S_i^z
\ee
where $\eta_i =\pm 1$ on the different sublattices, reflecting the type of order
expected at low temperatures, expands the partition function
\be
Z = \Tr \{ e^{-\beta H'} \} = \sum_{n=0}^{\infty} {(-\beta)^n \over n!} \Tr \{ (H-FM)^n\}
\ee
and compute the appropriate staggered suscetibility
\be
\chi = - {\partial^2\over \partial F^2} ( {1\over N} \ln Z )
\ee
in the form
\be
\chi = 1 + \sum_{n=1}^{\infty} a_n (\alpha) K^n \label{eq_chi}
\ee
where $K=\beta J_1$, $\alpha = J_2/J_1$ and a constant has been factored out.
Ref. [\onlinecite{pir66}] gives the $a_n(\alpha)$ for both AF$_1$ and AF$_2$ phases
through $n=6$. We have extended these series by 4 terms in the AF$_1$ phase and
3 terms in the AF$_2$ phase.
The coefficient are given in Table I.
Our results agree fully with previous shorter
series\cite{pir66,pan99} for the AF$_1$ phase, but there is a discrepancy at 6th order for
the AF$_2$ series. We believe that in the second last row of the last table in
Appendix II of Ref. [\onlinecite{pir66}] the entry 994944 should be replaced by -2230656.
With this substitution we find agreement for both $S=\half$ and $S=1$
(for which we have also derived short series).

The series have been analysed  via standard Pad\'e approximant methods\cite{gut}.
Table II shows the critical temperature and exponent estimates obtained from a direct analysis
of ${d\over dK} \ln \chi(K)$. As is apparent, for the smaller value of $\alpha=J_2/J_1$, there is a
consistent pole and an exponent around 1.4, which is the value expected for the
Heisenberg ($n=3$) universality class.
As $\alpha$ increases the series become less regular.
This is due to a pole on the negative real axis lying closer
to the origin (``ferromagnetic singularity"). Nevertheless the direct analysis is
consistent with $\gamma\simeq 1.4$ along the entire line. Assuming this, it is possible
to increase the precision in $K_c$. The resulting critical line is shown in Fig. 3.
In the AF$_2$ phase the series are less regular and consequently the analysis is less
precise. In Table III we show some of the raw analysis results. As is evident the higher-order
Pad\'es show a consistent physical singularity down to at least
$J_2/J_1=0.8$, although there is again an interfering singularity on
the negative real axis. The exponent estimates show a clear decreasing
trend from $J_1/J_2=0$ (while corresponds to the N\'eel phase of the simple
cubic lattice, for which we expect a second-order transition with the
universal exponent 1.40) for increasing $J_1/J_2$. For example, for
$J_1/J_2=1.25$ ($J_2/J_1=0.8$), the estimated $\gamma<1$. Now, if the AF$_2$-paramagnetic phase boundary
is second-order, we would expect critical exponents appropriate to an
$n=6$ component order parameter. Thus the series will be affected by a crossover from
$n=3$ to $n=6$, presumably followed by a second crossover to the bicritical point.
In a relatively short series this can give rise to varying critical exponents, as observed
here. On the other hand there are arguments, mentioned previously, for this
transition to be first-order. If that is the case then the transition will lie above our curve
in Fig. 4, and the apparent divergence locates a spinodal line within the order phase. The
``bicritical point" is then, in fact, a ``critical-end point"\cite{ban79}.


\section{Conclusions\label{sec4}}
We have used a combination of series expansions at $T=0$ and conventional
high-temperature series to elucidate the phase diagram of the quantum $S=\half$
Heisenberg antiferromagnet on the BCC lattice with nearest and next-nearest-neighbor
interactions. The ground state properties of this model have not, to
our knowledge, been  previously studied by series methods. We identify a
first-order quantum phase transition at $J_2/J_1\simeq 0.705\pm 0.005$
between the two possible types of antiferromagnetic order. We have extended previous
high temperature series by 4 and 3 terms, respectively in the AF$_1$ and AF$_2$ phases.
These longer series allow a fairly precise estimate of the AF$_1$-paramagnetic
critical line, and confirm $n=3$ universality along this
line. The nature of the AF$_2$-paramagnetic line remains enigmatic. We find
a consistent line of poles, which may represent a true critical line.
However we cannot exclude the possibility of a first-order transition at
somewhat higher temperatures.

In the corresponding Ising case the AF$_2$-paramagnetic transition was shown
to be first-order by a careful analysis using both high and low temperature
series\cite{vel83}. The former approach cannot be used here since low temperature
series cannot be obtained. It would be of some interest to study this system
using quantum Monte Carlo methods, although a ``sign problem" would be
expected.
A thorough Monte Carlo study for the classical Heisenberg system would also appear very
worthwhile. To our knowledge no such studies have been attempted.

\begin{acknowledgments}
This work  forms part of
a research project supported by a grant
from the Australian Research Council.
The computations were performed on an AlphaServer SC
 computer. We are grateful for the computing resources provided
 by the Australian Partnership for Advanced Computing (APAC)
National Facility.
\end{acknowledgments}

\newpage
\bibliography{basename of .bib file}

\begin{references}
\bibitem{sma66}J.S. Smart, ``Effective Field Theories of Magnetism", (Saunders, Philadelphia, 1966).
\bibitem{cha88}P. Chandra and B. Doucot, Phys. Rev. B {\bf 38}, 9335(1988).
\bibitem{cap00}L. Capriotti and S. Sorella, Phys. Rev. Lett. {\bf 84}, 3173(2000);
M.S.L. du Croo de Jongh, J.M.J. van Leeuwen, and W. van Saarloos, Phys. Rev. B{\bf 62}, 14844(2000).
\bibitem{sus01}O.P. Sushkov, J. Oitmaa and W. Zheng, Phys. Rev. B{\bf 63}, 104420(2001).
\bibitem{sch02}R. Schmidt, J. Schulenburg, J. Richter and D.D. Betts,
Phys. Rev. B{\bf 66}, 224406(2002).
\bibitem{pir66}K. Pirnie, P.J. Wood and J. Eve, Molecular Phys. {\bf 11}, 551(1966);
see also G.S. Rushbrooke, G.A. Baker Jr., and P.J. Wood, ``Phase Transitions and
Critical Phenomena", Vol. 3 Ch 5 ed C. Domb and M.S. Green (Academic, New York, 1974).
\bibitem{pan99}K.K. Pan, Phys. Rev. B{\bf 59}, 1168(1999).
\bibitem{ban79}J.R. Banavar, D. Jasnow and D.P. Landau, Phys. Rev. B {\bf 20}, 3820(1979).
\bibitem{vel83}M.J. Velgakis and M. Ferer, Phys. Rev. B{\bf 270}, 401(1983).

\bibitem{shn80}Y. Shnidman and D. Mukamel, J. Phys. C{\bf 13}, 5197(1980).
\bibitem{gel00} M.~P.~ Gelfand and R.~R.~P.~ Singh, Advances in Physics
{\bf 49}, 93 (2000).
\bibitem{zhe91}W. Zheng, J. Oitmaa and C.J. Hamer, Phys. Rev. B{\bf 43}, 8321(1991).
\bibitem{oit94}J. Oitmaa, C.J. Hamer and W. Zheng, Phys. Rev. B{\bf 50}, 3877(1994).
\bibitem{gut}A.J. Guttmann, in ``Phase Transitions and Critical
Phenomena'', Vol. 13 ed. C. Domb and J. Lebowitz (New York, Academic, 1989).

\end{references}


\begin{table*}
\squeezetable
\caption{Susceptibility polynomials $a_n(\alpha)$ (Eq. \ref{eq_chi}) for the spin-$\half$
bcc(1,2) Heisenberg antiferromagnet. To avoid fractions, the values of $2^{n+1} (n+1)! a_n(\alpha)$
are given.
}\label{tab_ser2}
\begin{ruledtabular}
\begin{tabular}{|rl|}
\multicolumn{2}{|c|}{AF$_1$ (N\'eel) phase}\\
  n=1  &             16,              -12 \\
  n=2  &            160,             -288,               72\\
  n=3  &           2048,            -5424,             4032,             -528 \\
  n=4  &          31096,          -110640,           132168,           -55680,             4950\\
  n=5  &         557456,         -2470992,          4070976,         -2947488,           825120,
      -56196\\
  n=6  &       11495460,        -61270752,        127951356,       -131155632,         65324256,
          -13458816,           693273\\
  n=7  &      269007424,      -1670678976,       4241825472,      -5589739344,       4034541024,
      -1489565472,        238760256,         -9630816 \\
  n=8  &     7026835032,     -49911466560,     148837552176,    -240765473760,     226149075384,
     -123139222944,      35480202336,  \\
     &    -4568142528,        156881934\\
  n=9  &   202835216096,   -1620119828784,    5556372919872,  -10631250963456,   12358650036192,
     -8857806164208,  \\
     &  3806668562496,    -888255818640,      94585942080,      -2810097960\\
 n=10  &  6406925312668,  -56883245890656,  220319075124360, -487248245389392,  674224356164040,
     -602298324975312,\\
     &  344178723975168, -120654960988368,   23412924414240,   -2116355382240,
         52557775149 \\
 \hline
\multicolumn{2}{|c|}{AF$_2$ phase} \\
  n=1   &           0,             12\\
  n=2   &         -40,              0,             84\\
  n=3   &         -64,           -240,              0,            720\\
  n=4   &        1320,           -528,          -5112,              0,           7422\\
  n=5   &        9968,           3408,         -11472,         -50784,              0,
            92412\\
  n=6   &     -131916,          13200,         197604,        -133248,       -1262928,
                0,        1323657\\
  n=7  &     -2026752,        6866304,        2195328,       12122160,       -3610560,
       -18223392,              0,       21144864\\
  n=8  &       556840,       68279328,     -335099184,       70461024,       -2954040,
       -61043232,     -432565248,              0,      375695526\\
  n=9  &    359570256,         205680,    -3444654528,    10167197376,      740209344,
      8510348784,    -1552929696,    -7905850224,              0,     7456498848\\
\end{tabular}
\end{ruledtabular}
\end{table*}

\begin{table}
\caption{Estimates of the critical temperature $K_c$  and
exponent $\gamma$ (in brackets) from Dlog Pad\'e approximants to the
N\'eel phase staggered susceptibility series.
}\label{tab2}
\begin{ruledtabular}
\begin{tabular}{|ccccc|}
$J_2/J_1$   & 0.0            & 0.2            & 0.4           & 0.6      \\
$[3/4]$     & 0.7233(1.391)  & 0.8881(1.427)  & 1.1614(1.418) & 1.6710(1.117)   \\
$[4/4]$     & 0.7248(1.410)  & 0.8839(1.391)  & 1.1634(1.429) & 1.7446(1.314)  \\
$[3/5]$     & 0.7251(1.414)  & 0.8844(1.396)  & 1.1635(1.430) & 1.7706(1.416)  \\
$[5/4]$     & 0.7281(1.459)  & 0.8896(1.454)  & 1.1704(1.478) &  $-$    \\
$[4/5]$     & 0.7199(1.371)  & 0.8859(1.410)  &  1.1606(1.415)& 1.6534(1.087) \\
negative pole & -1.0         & -0.8           & -0.68         &  -0.6 \\
\end{tabular}
\end{ruledtabular}
\end{table}

\begin{table}
\caption{Estimates of the critical temperature $K_c=J_2/kT_c$  and
exponent $\gamma$ (in brackets) from Dlog Pad\'e approximants to the
AF$_2$ phase staggered susceptibility series.
}\label{tab3}
\begin{ruledtabular}
\begin{tabular}{|ccccc|}
$J_1/J_2$   & 0.0           & 0.5            & 1.0    & 1.25      \\
$[3/3]$     & 1.0218(1.232) & 1.0698(1.299)  & $-$    & $-$ \\
$[2/4]$     & 1.0318(1.287) &  1.0709(1.304) &  $-$   & $-$ \\
$[3/4]$     & 1.0540(1.404) & 1.0847(1.367)  & 1.221(1.232) & 1.387(0.966) \\
$[4/4]$     & 1.0627(1.461) & 1.0827(1.357)  & 1.233(1.282) & 1.369(0.920)  \\
$[3/5]$     & 1.0639(1.472) & 1.0828(1.358)  &  1.234(1.288) & 1.370(0.924) \\
negative pole & -1.5 & -0.7 & -0.5 &  -0.4 \\
\end{tabular}
\end{ruledtabular}
\end{table}

\end{document}